\documentclass[12pt]{iopart}
\usepackage{graphicx}

\newcommand{\beq}{\begin{equation}}
\newcommand{\eeq}{\end{equation}}
\newcommand{\beqn}{\begin{eqnarray}}
\newcommand{\eeqn}{\end{eqnarray}}

\begin{document}

\title[General relativistic leakage scheme]
{An implementation of the microphysics in full general
relativity : General relativistic neutrino leakage scheme}

\author{Yuichiro Sekiguchi}

\address{Division of Theoretical Astronomy,
National Astronomical Observatory of Japan, Mitaka, Tokyo 181-8588,
Japan }
\ead{sekig@th.nao.ac.jp}

\begin{abstract}

Performing fully general relativistic simulations taking account of
microphysical processes (e.g., weak interactions and neutrino cooling)
is one of long standing problems in numerical relativity.
One of main difficulties in implementation of weak interactions in 
the general relativistic framework lies on the fact that the 
characteristic timescale of weak interaction processes 
(the WP timescale, $t_{\rm wp} \sim \vert Y_{e}/\dot{Y}_{e} \vert$) 
in hot dense matters is much shorter than the dynamical timescale 
($t_{\rm dyn}$).
Numerically this means that {\it stiff} source terms appears 
in the equations so that an implicit scheme is in general necessary to
stably solve the relevant equations. Otherwise a very short timestep 
($\Delta t$ $<$ $t_{\rm wp} \ll t_{\rm dyn}$) will be required to 
solve them explicitly, which is unrealistic in the present computational
resources. 
Furthermore, in the relativistic framework, 
the Lorentz factor is coupled with the rest mass density and the energy 
density. The specific enthalpy is also coupled with the momentum. 
Due to these couplings, it is very complicated to recover the primitive 
variables and the Lorentz factor from conserved quantities. 
Consequently, it is very difficult to solve the equations implicitly 
in the fully general relativistic framework. At the current status, no 
implicit procedure have been proposed except for the case of the 
spherical symmetry. 
Therefore, an approximate, explicit procedure is developed 
in the fully general relativistic framework
in this paper as an first implementation of the microphysics toward a 
more realistic sophisticated model.
The procedure is based on the so-called neutrino leakage schemes which 
is based on the property that the characteristic timescale in which 
neutrinos leak out of the system 
(the leakage timescale, $t_{\rm leak}$) is much longer than 
the WP timescale. In the previous leakage 
schemes, however, the problems of the stiff source terms are avoided in 
an artificial manner.
In this paper, I present a detailed neutrino leakage scheme and a simple 
and stable method for solving the equations explicitly in the fully 
general relativistic framework. The drawback of the artificial treatment 
of the stiff source terms is improved.
I also perform a test simulation to check the validity of the present 
method, showing that it works fairly well.

\end{abstract}

\maketitle

%%%%%%%%%%%%%%%%%%%%%%
\section{Introduction}
%%%%%%%%%%%%%%%%%%%%%%

\subsection{Motivation}

Numerical relativity is the unique and powerful tool to explore 
dynamical phenomena in which strong gravity plays important roles.
Stellar core collapse and mergers of compact star binaries are among 
the most important and interesting events in the field.
In theoretical view points, performing simulations of these phenomena
is one of challenging problems because a rich diversity of physics
has to be taken into account.
All four known forces of nature are involved and play important roles
during the collapse. General relativistic gravity plays essential roles
in formation of a black hole. Note also that general relativity may play 
important roles in supernova explosion as previous pioneering works
\cite{vanRiper,Takahara} showed. The weak interactions and emission of 
neutrinos govern energy and lepton-number losses, and hence 
driving the thermal and chemical evolutions of the system. 
The strong interactions determine ingredients and properties of
dense matters. Strong magnetic fields, if they are present, may modify
the dynamics.

There is a long list of studies which explore these phenomena in 
the framework of numerical relativity (see \cite{SCC} and references 
therein for resent simulations of stellar core collapse, and 
see \cite{BNS1,BNS2} and references therein for those of compact 
binary merger). 
In most of the previous studies, however, treatments of microphysics are 
very simplified and more sophisticated studies are necessary.

Furthermore, recent observations \cite{LGRBobs1-1,LGRBobs1-2,LGRBobs2-1,
LGRBobs2-2,LGRBobs2-3,LGRBobs3-1,LGRBobs3-2} have discovered the spectroscopic
connections between several supernovae and long gamma-ray bursts (GRBs), 
clarifying that at least some of long GRBs are associated with the collapse 
of massive stars. Also, there are theoretical models that a short GRB
occurs as a result of a binary neutron star merger 
\cite{SGRBmodel1,SGRBmodel2}.
The relevant process of the energy deposition to form a GRB fireball may be
pair annihilation of neutrinos emitted from a hot massive disk around 
a black hole formed after the collapse or the merger.
These also enhance the importance of exploring stellar core collapse 
and coalescence of compact star binary in full general relativity taking
account of microphysical processes.

Gravitational wave astronomy will start in this decade.
The first generation of ground-based interferometric detectors 
(LIGO \cite{LIGO}, VIRGO \cite{VIRGO}, GEO600 \cite{GEO600}) 
are now in the scientific search for gravitational waves. 
To obtain physically valuable information from these observations,
it is necessary to connect the observed data and the physics behind it.
For this purpose, performing numerical simulation is the unique approach.
However, accurate predictions of gravitational waveforms are still 
hampered by the facts that reliable estimates of waveforms require a general 
relativistic treatment \cite{Dimm02, SS03}, and that appropriate treatments 
of microphysics such as a nuclear equation of state (EOS), the electron 
capture, and neutrino emissions and transfers. 
General relativistic simulations including microphysics are required to make
accurate predictions of gravitational waveforms.

As described above, to perform multidimensional simulations in the frame
work of numerical relativity implementing microphysics is currently one of the
most important subjects in theoretical astrophysics. 
In spherical symmetry, fully general relativistic simulations 
\cite{Yamada99,Lieb04,Sumi05} of stellar core collapse have been performed 
in so-called state-of-the-art manners, namely, employing realistic 
equations of state, taking account of relevant microphysics, and 
solving the Boltzmann equation for the transfer of neutrinos (see also \cite{MM89}). 
In the multidimensional case, by contrast, there are few studies 
in the framework of general relativity. Recently,general relativistic simulations 
implementing a realistic EOS and the electron capture were performed 
\cite{Ott07,Dimm07}.
In their calculation, however, the electron capture rate is not
calculated in a self-consistent manner. Instead, they adopted
a simplified prescription proposed in \cite{LiebS}, which is 
based on results of spherically symmetric simulations.
It is not clear whether this treatment is justified for
non-spherical collapse and the multidimensional phenomena.
More importantly, they did not take account of neutrino cooling.

Recently, I have made a fully general relativistic code with
microphysics for the first time \cite{PhD}. 
Since it is currently impossible to fully solve the multidimensional 
neutrino transfer equations in the framework of full general 
relativity because of restrictions of computational resources,
it will be reasonable to adopt an approximated treatment of neutrino
cooling. In that work, a general relativistic version of the so-called 
neutrino leakage schemes is developed.

\subsection{Neutrino leakage scheme}

The leakage schemes \cite{leak1,leak2,leak5,leak3,leak4} as an
approximate method for the neutrino cooling has a well-established 
history (e.g. \cite{leak3}). The basic concept of the original neutrino 
leakage schemes \cite{leak1,leak2} is to treat the following two regions 
in the system separately: one region is where the diffusion timescale 
of neutrinos is longer than the dynamical timescale, and hence neutrinos 
are 'trapped' (neutrino-trapped region); the other region is where the 
diffusion timescale is shorter than the dynamical timescale, and hence 
neutrinos stream out freely out of the system (free-streaming region). 
The idea of treating the diffusion region separately has been applied to 
more advanced methods for the neutrino transfer
(see e.g., \cite{Ott08} and references therein).

Then, electron neutrinos and anti-neutrinos in the neutrino-trapped
region are assumed to be in $\beta$-equilibrium state. 
The {\it net} local rates of lepton-number and energy exchange with matters 
are set to be zero in the neutrino-trapped region.
To treat diffusive 'leakage' of neutrinos out of the 
neutrino-trapped region,
phenomenological source terms based on the diffusion theory are 
introduced \cite{leak1,leak2}.
In the free-streaming region, on the other hand, it is assumed that 
neutrinos escape from the system without interacting with matter. 
Therefore, neutrinos carry the lepton number and the energy according to
the local weak interaction rates. 
The neutrino fractions are not solved in the original version of the 
leakage scheme: Only the total lepton fraction is necessary in the 
free-streaming region and the neutrino fractions are set to zero in 
the free-streaming region. Note that there is a sharp
discontinuity between the two regions. Consequently, thermodynamical 
quantities, in particular those of neutrinos and the electron fraction, 
are also discontinuous at the boundary. 

The {\it transfer} of neutrinos are not solved in the leakage
schemes. Therefore, they cannot treat {\it non-local} interactions
among the neutrinos and matters; for example, the so-called neutrino
heating \cite{BW85} and the neutrino pair annihilation cannot be treated
in the leakage scheme.
Nevertheless, I consider a detailed general relativistic leakage scheme
presented in this paper to be an important step towards more reliable 
and sophisticated models, since the simulated physical timescales in the 
case of compact binary mergers will be order of 10 ms and neutrino
transfer is expected to be unimportant \cite{RL03}, and since the
neutrino heating would be not very important in the case of prompt
black hole formation.

Usually, the boundary between the neutrino-trapped and free-streaming
regions is given by hand as a single 'neutrino-trapping' density 
($\rho_{\rm trap}$) in the previous simulations of stellar core collapse 
\cite{leak1,leak2,leak4}. In fact, however, the location at which the 
neutrino trapping occurs depends strongly on the neutrino energies 
($\epsilon_{\nu}$), and hence, there are different neutrino-trapping 
densities for different neutrino energies. 
The neutrino-trapping densities depend strongly on the neutrino energies 
as $\rho_{\rm trap} \propto \epsilon_{\nu}^{-3}$ \cite{Bethe1990}.
This implies that neutrinos with lowest energy leave their
corresponding neutrino-trapping region first, and neutrinos with higher energy
are emitted later.

In the previous leakage schemes \cite{leak1,leak2,leak4}, on the other
hand, all neutrinos are emitted in one moment irrespective of their energy.
Consequently in the case of the so-called neutrino burst
emission (e.g., \cite{Bethe1990}), for example, 
the duration in which the neutrinos
are emitted is shortened and the peak luminosity at the burst is
overestimated in the previous leakage schemes \cite{leak2,PhD}.
The dependence of the neutrino-trapping densities and neutrino diffusion rates 
on the neutrino energies are approximately taken into account 
in the recent simulations of binary neutron star mergers
\cite{RJS96,RL03}. 
However the lepton-number conservation equations for neutrinos are not
solved \cite{RJS96}.

Recently, a numerical code based on a relativistic extension of the
leakage schemes was developed in \cite{PhD}, where not the region of the system
but the energy momentum tensor of neutrinos are decomposed into two parts;
'trapped-neutrino' and 'streaming-neutrino' parts. However the source
terms of hydrodynamic and the lepton-number-conservation equations are 
determined using the single
neutrino-trapping density as in the case of the previous leakage schemes.
More recently, Liebend\"orfer et
al. \cite{Lieb09} proposed a scheme, which they call the isotropic
diffusion source approximation, where the neutrino distribution function 
is decomposed into an isotropic distribution function of trapped
neutrinos and a distribution function of streaming neutrinos.

The present work is based on the previous studies described above. 
The framework of general relativistic extension of leakage scheme is 
based on my previous study in \cite{PhD}. The treatment of neutrino diffusion
rates is based on the recent work by Rosswog and Liebend\"orfer \cite{RL03}
where the neutrino-energy dependences are taken into account.
Thus the remaining main problem to implement the relevant microphysics is 
that straightforward explicit scheme cannot be adopted to solve the
equations \cite{Bruenn85} since the characteristic timescale of weak
interaction processes ($t_{\rm wp} \sim \vert Y_{e}/\dot{Y}_{e} \vert$,
hereafter the WP timescale) is much shorter than the dynamical timescale 
($t_{\rm dyn}$) in hot dense regions, as described in
Sec. \ref{Difficulty}. Note that the WP timescale is different from the
so-called weak timescale.
In this paper I present a simple and stable method in which the
equations are solved explicitly in the {\it dynamical timescale} 
in the fully general relativistic framework.

The paper is organized as follows. 
First, main difficulties of implementation of weak interactions and 
neutrino cooling in full general relativity compared to implementation of 
them in Newtonian framework are briefly summarized in Sec. \ref{Difficulty}.
Then, framework of the implementation of the microphysics is described 
in detail in Sec.~\ref{GRleak}. 
Some details of microphysics and numerics are described in Sec.~\ref{Details},
although GR leakage framework is independent of specific implementations of 
microphysics. 
In Sec.~\ref{Results}, results of a test simulation is briefly described 
illustrating good ability of the present implementation.
Section \ref{Summary} is devoted to the summary and discussions.
Throughout the paper, the geometrical unit $c=G=1$ is used otherwise stated.

%%%%%%%%%%%%%%%%%%%%%%
\section{Difficulties of implementation of weak interactions and
 neutrino cooling in full general relativity}\label{Difficulty}
%%%%%%%%%%%%%%%%%%%%%%

Since the characteristic timescale of weak interaction precesses 
(the WP timescale $t_{\rm wp} \sim \vert Y_{e}/\dot{Y}_{e} \vert $) 
is much shorter than the dynamical timescale $t_{\rm dyn}$ in hot 
dense matters \cite{Bruenn85,RL03}, the numerical treatment of the 
weak interactions cannot be explicit,
as noted in the previous pioneering work by Bruenn \cite{Bruenn85}.
Otherwise a very short timestep 
($\Delta t$ $<$ $t_{\rm wp} \ll t_{\rm dyn}$) will be required to solve 
the equations explicitly, which is unrealistic in the present 
computational resources.

The {\it net} rates of lepton-number and energy exchanges between 
matters and neutrinos may not be large, and consequently an {\it effective} 
timescale apparently may not be short compared to the dynamical
timescale. However, this does not immediately imply that one can solve the
equations explicitly without bringing in any devices.
For example the net electron capture rate vanishes in the
$\beta$-equilibrium. 
The achievement of $\beta$-equilibrium is consequences of both 
cancellation of two very {\it large} weak interaction processes
(the electron and the electron-neutrino captures)
and the action of the phase space blocking.
Note that the weak interaction processes depend enormously on the 
temperature and the lepton chemical potentials.
Therefore, small error in evaluations of the temperature and a small
deviation from the $\beta$-equilibrium due to small error in estimation
of the lepton fractions will produce large error and stiff source terms, 
and consequently the explicit numerical evolutions may become unstable.

In the following of this section, I describe difficulties of
implementation of weak interactions and neutrino cooling into the 
hydrodynamic equations in the conservative schemes in full general
relativity {\it compared with the Newtonian framework}.

In the Newtonian framework, the equations may be solved implicitly
\cite{BW82,Bruenn85,MBHLSR87,MB93,RJ02,Livne04,Buras06,Burrows07,MJ09}
(see also \cite{MM99,Ott08} and references therein). 
The equations of hydrodynamics, lepton-number conservations, and neutrino
processes are schematically written as,
\beqn
\dot{\rho   } &=& 0 , \\
\dot{v_{i}  } &=& S_{v_{i}  }(\rho, Y_{e}, T, Q_{\nu}) , \\
\dot{Y_{e}  } &=& S_{Y_{e}  }(\rho, Y_{e}, T, Q_{\nu}) , \\
\dot{e      } &=& S_{e      }(\rho, Y_{e}, T, Q_{\nu}) , \\
\dot{Q_{\nu}} &=& S_{Q_{\nu}}(\rho, Y_{e}, T, Q_{\nu}) , 
\eeqn
where $\rho$ is the rest mass density, $v_{i}$ is the velocity, 
$Y_{e}$ is the electron fraction,
$e$ is the (internal) energy of matter, $T$ is the temperature and 
$Q_{\nu}$ stands for the relevant neutrino quantities. 
$S$'s in the right hand side stand for the relevant source terms. 
Comparing the quantities in the left-hand-side and the argument quantities 
in the source terms, only the relation between $e$ and $T$ is nontrivial.
Usually, EOSs employed in the simulation is tabularized, and 
one dimensional search over the EOS table is required to solve them.
To achieve this procedure in an implicit manner is  quite a difficult problem.

In the relativistic framework, the situation becomes much more
complicated in conservative schemes, since the Lorentz factor ($\Gamma$)
is coupled with rest mass density and the energy density (see
Eqs. (\ref{continuS}) and (\ref{eneS})), and since the specific enthalpy
($h = h(\rho,Y_{e},T)$) is coupled with the momentum (see
Eq. (\ref{momS})).

It should be addressed that the previous fully general relativistic
works in the spherical symmetry \cite{Yamada99,Lieb04} are based on 
the so-called Misner-Sharp coordinates \cite{MS64}.
There are no such complicated couplings in this Lagrangian coordinates.
Accordingly the equations may be solved essentially in the same
manner as in the Newtonian framework.
%Since no such simple Lagrangian coordinates are not known in the
%multidimensional case, the complicated couplings inevitably appear 
%in the conservative schemes.
In multidimensional case, on the other hand, no Lagrangian coordinates 
are known, and the Eulerian coordinates are adopted.
In the Eulerian coordinate, the complicated couplings inevitably appear 
in multidimensional case.

Omitting the factors associated with the geometric variables 
(which are known when solving hydrodynamics equations), 
the equations to be solved in the relativistic framework are 
schematically written as,
\beqn
\dot{\rho_{*}}(\rho,\Gamma)  &=& 0 , \label{rhoEq} \\
\dot{\hat{u}}_{i}(u_{i},h) = \dot{\hat{u}}_{i}(u_{i},\rho,Y_{e},T) 
        &=& S_{\hat{u}_{i}}(\rho, Y_{e}, T, Q_{\nu}, \Gamma) , \\
\dot{Y_{e}  } &=& S_{Y_{e}  }(\rho, Y_{e}, T, Q_{\nu}, \Gamma) , \\
\dot{\hat{e}}(\rho, Y_{e}, T, \Gamma) &=& 
S_{\hat{e}}(\rho, Y_{e}, T, Q_{\nu}, \Gamma) , \\
\dot{Q_{\nu}} &=& S_{Q_{\nu}}(\rho, Y_{e}, T, Q_{\nu}, \Gamma) , 
\eeqn
where $\rho_{*}$ is a weighted density, $\hat{u}_{\alpha}$ is a weighted 
four velocity, $\hat{e}$ is a weighted energy density (see Sec.~\ref{BasicEq} 
for the definition of these variables).
Note that the Lorentz factor is appeared in the equations.

The Lorentz factor is calculated by solving the normalization condition 
$u^{\alpha}u_{\alpha}=-1$, which is rather complicated nonlinear equation 
schematically written as
\beq
f_{\rm normalization}(\hat{u_{i}}, \Gamma) 
= f_{\rm normalization}(u_{i}, \rho, Y_{e}, T, \Gamma) = 0. \label{nomEq}
\eeq
(For a solution of the equation in practice, see Sec. \ref{Reconst}.)
The accurate calculation of the Lorentz factor and the accurate solution
of the normalization condition are very important in the numerical
relativistic hydrodynamics.

Now, it is obvious that the argument quantities in the source terms are
not simply related with the left-hand-side evolved quantities in Eqs. 
(\ref{rhoEq})--(\ref{nomEq}). To solve the equations implicitly is quite
difficult and there are even no successful formulations.
Moreover it is not clear whether a convergent solution can be {\it
stably} obtained
numerically or not, since they are simultaneous nonlinear equations.
Therefore, it may not be a poor choice to adopt an alternative approach 
in which the equations are solved explicitly with some approximations 
(see Sec. \ref{GRleak}).

The second minor problem which is not exist in the Newtonian framework 
is that one cannot add any microphysical processes in forms of 'cooling'
or 'heating' terms $Q_{\alpha}$, into the right hand side of hydrodynamic
equations as
\beq
\nabla_{\alpha}T^{\beta}_{\beta} = Q_{\beta}.
\eeq
Instead, the energy-momentum tensor of neutrinos should be introduced in
the general relativistic framework:
\beq
(T^{\rm tot})_{\alpha\beta} = 
(T^{{\rm F}})_{\alpha\beta} + (T^{\nu})_{\alpha\beta},
\eeq
where $T^{\rm tot}$, $T^{\rm F}$, and $T^{\nu}$ are 
the total energy-momentum tensor
and energy-momentum tensor of fluid and neutrino parts, respectively 
(see Sec. \ref{GRleak} for the definition). 
Now, one should solve the following coupled equations
\beqn
\nabla_{\alpha}(T^{{\rm F}})^{\alpha}_{\beta} &=& Q_{\beta}, \label{T_Eq1} \\
\nabla_{\alpha}(T^{\nu})^{\alpha}_{\beta} &=& -Q_{\beta} \label{T_Eq2}.
\eeqn
Here, the source term $Q_{\alpha}$ can be regarded as the local production 
of neutrinos through the weak processes, ignoring non-local neutrino 
capture on matter. 
%Note that an implicit scheme may be in general
%required to solve them since $Q_{\alpha}$ is characterized by the weak
%timescale.

%Shortly speaking, the above two problems mainly prevent one to perform 
%simulations taking account of the microphysical processes in fully
%general relativistic framework.

%%%%%%%%%%%%%%%%%%%%%%
\section{General relativistic neutrino leakage scheme}\label{GRleak}
%%%%%%%%%%%%%%%%%%%%%%

In the following, I describe in some detail a method for solving all 
of the equation in an explicit manner.
As described in the previous section, since $t_{\rm wp} \ll t_{\rm dyn}$ 
in the hot dense matter regions, the source terms in the equations 
become too {\it stiff} to be solved explicitly. 
The characteristic timescale of leakage of neutrinos
from the system $t_{\rm leak}$, however, is much longer 
than $t_{\rm wp}$ in the hot dense matter region. 
Note that $t_{\rm leak} \sim L/c \sim t_{\rm dyn}$ where $L$ is the
characteristic length scale of the system. On the other hand,
$t_{\rm leak}$ is comparable to $t_{\rm wp}$ in the free-streaming
regions where the WP timescale is longer than or comparable with 
the dynamical timescale. 
Utilizing these facts, I approximate the original matter equations and reformulate them 
so that the source terms are to be characterized by the leakage timescale 
$t_{\rm leak}$.

%%%%%%%%%%%
\subsection{Decomposition of neutrino energy-momentum tensor}\label{EnergyMomentum}

Now, the problem is that the source term $Q_{\alpha}$ in 
Eqs. (\ref{T_Eq1}) and (\ref{T_Eq2}) becomes too {\it stiff} to solve
explicitly in hot dense matter regions where 
$t_{\rm wp} \ll t_{\rm dyn}$.
To overcome the situation, the following procedures are adopted.

First, it is assumed that the energy-momentum tensor of neutrinos are
be decomposed into 
'trapped-neutrino' ($(T^{\nu,{\rm T}})_{\alpha\beta}$) 
and 'streaming-neutrino' ($(T^{\nu,{\rm S}})_{\alpha\beta}$) parts as
\cite{PhD},
\beq
(T^{\nu})_{\alpha\beta} = (T^{\nu,{\rm T}})_{\alpha\beta} +
                       (T^{\nu,{\rm S}})_{\alpha\beta} . 
\label{nudecompose}
\eeq
Here, the trapped-neutrinos phenomenologically represent neutrinos which 
interact sufficiently frequently with matter and are thermalized, 
while the streaming-neutrino part describes a phenomenological flow of 
neutrinos streaming out of the system \cite{PhD} (see also \cite{Lieb09}
where a more sophisticate method based on the distribution function 
is adopted in the Newtonian framework).

Second, the locally produced neutrinos are assumed 
to {\it leak out} to be the streaming-neutrinos
with a leakage rate $Q^{\rm leak}_{\alpha}$:
\beq
\nabla_{\beta}(T^{\nu,{\rm S}})^{\beta}_{\alpha} =  Q^{\rm leak}_{\alpha}.
\label{T_Eq_nuS}
\eeq
Then, the equation of the trapped-neutrino part becomes
\beq
\nabla_{\beta}(T^{\nu,{\rm T}})^{\beta}_{\alpha} = 
Q_{\alpha} - Q^{\rm leak}_{\alpha}.
\label{T_Eq_nuT}
\eeq

Third, the trapped-neutrino part is combined with the fluid part to give
\beq
T_{\alpha\beta} \equiv (T^{\rm F})_{\alpha\beta} 
+ (T^{\nu,{\rm T}})_{\alpha\beta},
\eeq
and Eqs. (\ref{T_Eq1}) and (\ref{T_Eq_nuT}) are combined to give
\beq
\nabla_{\beta}T^{\beta}_{\alpha} = -Q^{\rm leak}_{\alpha} \label{T_Eq_M}.
\eeq
Thus the equations to be solved is changed from 
Eqs. (\ref{T_Eq1}) and (\ref{T_Eq2}) to 
Eqs. (\ref{T_Eq_M}) and (\ref{T_Eq_nuS}).
Note that the new equations only include the source terms $Q^{\rm leak}_{\alpha}$ 
which is characterized by the leakage timescale $t_{\rm leak}$.
Definition of $Q^{\rm leak}_{\alpha}$ will be found in Sec. \ref{leakagerate}.

The energy-momentum tensor of the fluid and trapped-neutrino parts
($T_{\alpha \beta}$) is treated as that of the perfect fluid
\beq
T_{\alpha\beta} = (\rho + \rho \varepsilon + P)
 u_{\alpha}u_{\beta} + P g_{\alpha\beta}, \label{T_fluid}
\eeq
where $\rho$ and $u^{\alpha}$ are the rest mass density and the 4-velocity.
The specific internal energy density ($\varepsilon$) and the pressure ($P$) 
are the sum of the contributions from the baryons 
(free protons, free neutrons, $\alpha$-particles, and heavy nuclei), 
leptons (electrons, positrons, and {\it trapped-neutrinos}), and the radiation as,
\beqn
P &=& P_{B} + P_{e} + P_{\nu} + P_{r}, \\
\varepsilon &=&  
\varepsilon_{B} + \varepsilon_{e} +
\varepsilon_{\nu} + \varepsilon_{r} ,
\eeqn
where subscripts '$B$', '$e$', '$r$', and '$\nu$' denote the components
of the baryons, electrons and positrons, radiation, and trapped-neutrinos, 
respectively. 

The streaming-neutrino part, on the other hand,  is set to be a general
form of
\beq
(T^{\nu,{\rm S}})_{\alpha\beta}= 
E n_{\alpha}n_{\beta} + F_{\alpha}n_{\beta} + F_{\beta}n_{\alpha} + P_{\alpha\beta},
\label{T_neutrino}
\eeq
where $F_{\alpha}n^{\alpha}=P_{\alpha \beta}n^{\alpha}=0$. 
In order to close the system,
we need an explicit expression of $P_{\alpha \beta}$.
In this paper, I adopt a rather simple form 
$P_{\alpha \beta}=\chi E \gamma_{\alpha \beta}$ with $\chi = 1/3$.
This approximation may work well in high density regions but
will violate in low density regions. However, the violation will 
not affect the dynamics since the total amount of 
streaming-neutrinos emitted in low density regions will be small.
Of course, a more sophisticated treatment will be necessary in a future
study.

%%%%%%%%%%%
\subsection{The lepton-number conservation equations}\label{Lepton}

The conservation equations of the lepton fractions can be written schematically as
\beqn
&&\!\! \frac{d Y_{e}}{dt} = -\gamma_{e} , \label{dYe} \\
&&\!\! \frac{d Y_{\nu e}}{dt} = \gamma_{\nu e},  \label{dYnu} \\ 
&&\!\! \frac{d Y_{\bar{\nu} e}}{dt} = \gamma_{\bar{\nu} e},  \label{dYna} \\ 
&&\!\! \frac{d Y_{\nu x}}{dt} = \gamma_{\nu x},  \label{dYno}  
\eeqn
where $Y_{e}$, $Y_{\nu e}$, $Y_{\bar{\nu} e}$, and $Y_{\nu x}$ denote
the electron fraction, the electron neutrino fraction, the electron
anti-neutrino fraction, and $\mu$ and $\tau$ neutrino and anti-neutrino 
fractions, respectively. 
It should be addressed that, in the previous simulations based on the 
leakage schemes \cite{leak1,leak2,leak4,RJS96}, the neutrino fractions 
are not solved.

The source terms of neutrino fractions can be written, on the basis of 
the present leakage scheme, as 
\beqn
&&\!\! \gamma_{\nu e} = \gamma_{\nu e}^{\rm local} - \gamma_{\nu e}^{\rm leak}, \\
&&\!\! \gamma_{\bar{\nu} e} = \gamma_{\bar{\nu} e}^{\rm local} 
                        - \gamma_{\bar{\nu} e}^{\rm leak}, \\
&&\!\! \gamma_{\nu x} = \gamma_{\nu x}^{\rm local} - \gamma_{\nu x}^{\rm leak}, 
\eeqn
where $\gamma_{\nu}^{\rm local}$ and $\gamma_{\nu}^{\rm leak}$ are
the local production and the leakage rates of neutrinos, respectively
(see Sec. \ref{leakagerate}).
Note that only the trapped-neutrinos are responsible for 
the neutrino fractions.
The thermodynamical quantities (e.g., the pressure and the chemical
potentials) of neutrinos can be calculated from the neutrino fractions
on the assumption of thermalization of the trapped neutrinos. 

The source term for the electron fraction conservation can be written
\beq
\gamma_{e} = \gamma_{\nu e}^{\rm local} - \gamma_{\bar{\nu} e}^{\rm local}.
\eeq
Since $\gamma^{\rm local}_{\nu}$\,s are characterized by
by the WP timescale $t_{\rm wp}$, some procedures are necessary to solve
the lepton conservation equations explicitly.
The following simple procedures are proposed to solve the equation stably. 

First, in each timestep $n$, the conservation equation of 
the {\it total} lepton fraction ($Y_{l}=Y_{e}-Y_{\nu e}+Y_{\bar{\nu} e}$), 
\beqn
&&\!\! \frac{d Y_{l}}{dt} = -\gamma_{l},  \label{dYl} 
\eeqn
is solved together with the conservation equation of $Y_{\nu x}$, Eq. (\ref{dYno}),
in advance of solving whole of the lepton conservation 
equations (Eqs. (\ref{dYe}) -- (\ref{dYno})).
Note that the source term 
$\gamma_{l} = \gamma_{\nu e}^{\rm leak} - \gamma_{\bar{\nu} e}^{\rm leak}$ 
is characterized by the leakage timescale $t_{\rm leak}$
so that this equation may be solved explicitly in the hydrodynamic timescale.
Then, assuming that the $\beta$-equilibrium is achieved, 
values of the lepton fractions in the $\beta$-equilibrium ($Y_{e}^{\beta}$,
$Y_{\nu e}^{\beta}$, and $Y_{\bar{\nu} e}^{\beta}$) are calculated from
evolved $Y_{l}$. 

Second, regarding $Y_{\nu e}^{\beta}$ and $Y_{\bar{\nu} e}^{\beta}$ as the
maximum allowed values of the neutrino fractions in the next 
timestep $n+1$,  the source terms are limited so that $Y_{\nu}$'s in 
the timestep $n+1$ do not exceed $Y_{\nu}^{\beta}$'s.
Then, the whole of the lepton conservation equations 
(Eqs. (\ref{dYe}) -- (\ref{dYno})) are solved explicitly utilizing the
limiters.

Third, the following conditions are checked 
\beqn
\mu_{p}+\mu_{e} < \mu_{n}+\mu_{\nu e} , \\
\mu_{n}-\mu_{e} < \mu_{p}+\mu_{\bar{\nu} e},
\eeqn
where $\mu_{p}$, $\mu_{n}$, $\mu_{e}$, $\mu_{\nu e}$ and $\mu_{\bar{\nu}
e}$ are the chemical potentials of protons, neutrons, electrons, 
electron neutrinos, and electron anti-neutrinos, respectively. 
If both conditions are satisfied, the values of
the lepton fractions in the timestep $n+1$ is set to be those in 
the $\beta$-equilibrium value;  
$Y_{e}^{\beta}$, $Y_{\nu e}^{\beta}$, and $Y_{\bar{\nu} e}^{\beta}$.
On the other hand, if either or both conditions are not satisfied, 
the lepton fractions in the timestep $n+1$ is set to be those obtained
by solving whole of the lepton-number conservation equations.  

A limiter for the evolution of $Y_{\nu x}$ may be also necessary 
in some case where the pair processes are dominant, for example, 
in simulations of collapse of population III stellar core.
In this case, the value of $Y_{\nu x}$ at the pair equilibrium 
(i.e. at $\mu_{\nu x}=0$), $Y_{\nu x}^{\rm pair}$ may be used 
to limit the source term.

In the present implementation it is not necessary to somewhat
artificially divide the system into neutrino-trapped and 
free-streaming regions. Therefore there is no discontinuous boundary
which existed in the previous leakage schemes \cite{leak1,leak2,leak4}.

I found that simulations of the collapse of population III stellar core
and the formation of a black hole, in which very high temperatures ($T>100$
MeV) are achieved, can be stably performed using the simple procedure
presented in this paper.

%%%%%%%%%%%
\subsection{Definition of leakage rates}\label{leakagerate}

In this subsection the definitions of the leakage rates $Q_{\alpha}^{\rm
leak}$ and $\gamma_{\nu}^{\rm leak}$ are presented.
Because $Q^{\rm leak}_{\nu}$ may be regarded as the emissivity of
neutrinos measured in the {\it fluid rest frame}, $Q^{\rm
leak}_{\alpha}$ is defined as \cite{SSR07}
\beq
Q^{\rm leak}_{\alpha} = Q^{\rm leak}_{\nu}u_{\alpha}.
\eeq\label{leakage_source_Q}
Note that although there may be a freedom to include terms $H_{\alpha}$
which satisfies $H_{\alpha}u^{\alpha} = 0$, Eq. (\ref{leakage_source_Q})
may be the best choice in the present framework.

The leakage rates $Q^{\rm leak}_{\nu}$ and 
$\gamma^{\rm leak}_{\nu}$ are assumed to satisfy the following properties.
\begin{enumerate}
\item The leakage rates approach the local rates 
  $Q_{\nu}^{\rm local}$ and $\gamma_{\nu}^{\rm local}$ in the low density, 
  transparent region. 
\item The leakage rates approach the diffusion rates 
  $Q_{\nu}^{\rm diff}$ and $\gamma_{\nu}^{\rm diff}$ in the high density,
  opaque region. 
\item The above two limits should be connected smoothly.
\end{enumerate}
Here, the local rates can be calculated based on the theory of weak 
interactions (see Sec. \ref{Local} for the local rates adopted in this paper) 
and the diffusion rates can be determined based on the 
diffusion theory (see Sec. \ref{Diff} for the definition of the diffusion rate
adopted in this paper). 
There will be several prescriptions to satisfy the requirement (iii)
\cite{RJS96,RL03}.
In this paper, the leakage rates are defined as
\beqn
&&\!\! Q_{\nu}^{\rm leak}= (1-e^{-b\tau_{\nu}}) Q_{\nu}^{\rm diff} 
+ e^{-b\tau_{\nu}} Q_{\nu}^{\rm local}, \label{Q_leak} \\
&&\!\! \gamma_{\nu}^{\rm leak}= (1-e^{-b\tau_{\nu}}) \gamma_{\nu}^{\rm diff} 
+ e^{-b\tau_{\nu}} \gamma_{\nu}^{\rm local}, 
\eeqn
where $\tau_{\nu}$ is the optical depth of neutrinos and $b$ is a parameter
which is typically set as $b^{-1}=2/3$. The optical depth can be
computed from the cross sections in a standard manner \cite{RJS96,RL03}.

%%%%%%%%%%%
\subsection{Explicit forms of basic equations in leakage scheme}\label{BasicEq}

The basic equations for the general relativistic hydrodynamics are the
continuity equation, the lepton-number conservation equations, and
the local conservation equation of the
energy-momentum. The explicit forms of the equations are presented in
this subsection for the purpose of convenience.

\subsubsection{The Continuity and lepton-number conservation equations}

The continuity equation is
\beq
\nabla_{\alpha}(\rho u^{\alpha}) = 0 \label{conti}.
\eeq
As fundamental variables for numerical simulations, the following
quantities are introduced:
$\rho_{\ast} \equiv \rho w e^{6\phi}$ and $v^{i} \equiv \frac{u^{i}}{u^{t}}$
where $ w \equiv \alpha u^{t}$.
Then, the continuity equation is written as
\beq
 \partial_{t}(\rho_{\ast} \sqrt{\eta}) 
 + \partial_{k}(\rho_{\ast}v^{k} \sqrt{\eta}) = 0 ,
\label{continuS} 
\eeq
where $\sqrt{\eta} \equiv \sqrt{\det \eta_{ij}}$ is the volume element
of the flat space in the curvilinear coordinates.

The lepton-number conservation equations (\ref{dYe}) -- (\ref{dYno}) 
can be abbreviated as
\beq
\frac{d Y_{L}}{dt} = \gamma_{L},
\eeq
Using the continuity equation, they become
\beq
\partial_{t}(\rho_{\ast}Y_{L} \sqrt{\eta}) 
+ \partial_{k}(\rho_{\ast}Y_{L}v^{k} \sqrt{\eta}) = \rho_{*}\gamma_{L}.
\label{e-Y} 
\eeq

\subsubsection{Energy-momentum conservation}

As discussed in Sec. \ref{EnergyMomentum}, I solve the following equations.
\beqn
&&\!\! \nabla_{\beta}T^{\beta}_{\alpha} = -Q^{\rm leak}_{\alpha} , \\
&&\!\! \nabla_{\beta}(T^{\nu,{\rm S}})^{\beta}_{\alpha} =  Q^{\rm leak}_{\alpha},
\eeqn
where $T_{\alpha \beta}$ and $(T^{\nu,{\rm S}})_{\alpha \beta}$ are given by
Eqs. (\ref{T_fluid}) and (\ref{T_neutrino}), respectively.
The source term $Q_{\alpha}^{\rm leak}$ is defined by Eq. (\ref{Q_leak}).

As fundamental variables for numerical simulations, I define the 
quantities
$\hat{u}_{i} \equiv hu_{i}$ and 
$\hat{e} \equiv hw - P(\rho w)^{-1}$.
Then, the Euler
equation ($\gamma_{i}^{\alpha} \nabla_{\beta}
T^{\beta}_{\ \alpha} = \gamma_{i}^{\alpha} Q_{\alpha}$), 
and the energy equation 
($n^{\alpha}\nabla_{\beta}T^{\alpha}_{\beta}=n^{\alpha}Q_{\alpha}$) can be 
written as
\beqn
&&\!\! \partial_{t}(\rho_{\ast} \hat{u}_{A} \sqrt{\eta})
+ \partial_{k}\left[\left\{
               \rho_{\ast} \hat{u}_{A} v^{k}
          + P\alpha e^{6\phi}\delta^{k}_{\ A} \right\} \sqrt{\eta}
      \right]  
\nonumber \\
&&\!\!\!\!  \ \ \ \ \ \ \ \ \ \ 
=  
- \rho_{\ast}\left[ 
w h \partial_{A}\alpha - \hat{u}_{i}\partial_{A}\beta^{i}
+ \frac{\alpha
 e^{-4\phi}}{2wh}\hat{u}_{k}\hat{u}_{l}\partial_{A}\tilde{\gamma}^{kl}
- \frac{2\alpha h (w^{2} -1)}{w}\partial_{A}\phi \;
\right] \nonumber \\ 
&&\!\!\!\!  \ \ \ \ \ \ \ \ \ \ \ \ \ 
 + P\partial_{A}(\alpha e^{6\phi}) 
 +\frac{P\alpha e^{6\phi} \delta^{\varpi}_{A}}{\varpi}  
 + \alpha e^{6\phi} Q_{A}, \label{momS}  \\
%%%
&&\!\!\!\! \partial_{t}\left( \rho_{\ast}\hat{u}_{\varphi} \sqrt{\eta} \right)
 + \partial_{k}\left( \rho_{\ast}\hat{u}_{\varphi}v^{k}  \sqrt{\eta}\right)
=    \alpha e^{6\phi} Q_{\varphi}, \\
%%%
&&\!\!\!\! \partial_{t}(\rho_{\ast}\hat{e} \sqrt{\eta}) + 
  + \partial_{k}\left[ \left\{
      \rho_{\ast}v^{k}\hat{e}
    + P e^{6\phi}\sqrt{\eta}(v^{k}+\beta^{k})  \right\} \sqrt{\eta}
  \right]
\nonumber \\
&&\!\!\!\!  \ \ \ \ \ \ \ \ \ \ \ 
= \alpha e^{6\phi} \sqrt{\eta}PK + 
\frac{\rho_{\ast}}{u^{t}h} \hat{u}_{k}\hat{u}_{l}K^{kl}
 - \rho_{\ast}\hat{u}_{i}\gamma^{ij}D_{j}\alpha 
 + \alpha e^{6\phi} Q_{\alpha}n^{\alpha},  \label{eneS} 
\eeqn
where the subscript $A$ denotes $\varpi$ or $z$ component.

The evolution equations of streaming-neutrinos 
($E$ and $F_{i}$) are written as
\beqn
&&\!\!\!\!\!\!\!\!
     \partial_{t}(\sqrt{\gamma}E) 
  + \partial_{k}\left[\sqrt{\gamma}(\alpha F^{k} - \beta^{k}E)\right]  
  = \sqrt{\gamma}\left(
      \alpha P^{kl}K_{kl} - F^{k}\partial_{k}\alpha + \alpha Q^{\rm leak}_{a}n^{a} 
    \right), \\
&&\!\!\!\!\!\!\!\!
    \partial_{t}(\sqrt{\gamma}F_{i}) 
  + \partial_{k}\left[\sqrt{\gamma}(\alpha P^{k}_{i}-\beta^{k}F_{i})\right]
  \nonumber \\
&&\!\!  \ \ \ \ \ \ \ \ \ \ \ \ \ \ \ \ \ \ \ \ \ \ \ \ \ \ \ 
  = \sqrt{\gamma}\left(
    -E\partial_{i}\alpha + F_{k}\partial_{i}\beta^{k} 
    + \frac{\alpha}{2}P^{kl}\partial_{i}\gamma_{kl} + \alpha Q^{\rm leak}_{i}
    \right).
\eeqn

%%%%%%%%%%%
\subsection{Recover of primitive variables} \label{Reconst}

In each numerical timestep, the so-called primitive variables
($\rho$, $Y_{L}$, $T$, and $v_{i}$) and the Lorentz factor 
$w=\alpha u^{t}=\sqrt{1+\gamma^{ij}u_{i}u_{j}}$ must be calculated from 
the conserved quantities
($\rho_{*}$, $\rho_{*}Y_{L}$, $\hat{e}$, and  $\hat{u}_{i}$), where
$Y_{L}$ is the representative of the lepton fractions.
Since the equation of state (EOS) of the nuclear matter are usually
tabularized in terms of the argument quantities 
($\rho$, $Y_{p} (=Y_{e})$, $T$),
I am devoted to the cases of the tabularized EOS in the following.

In the case where the whole of the lepton-number conservation equations are 
solved (see Sec. \ref{Lepton}),
the argument quantities ($\rho$, $Y_{e}$, $T$) are calculated
from the conserved quantities as follows.
\begin{enumerate}
\item Give a trial value, $\tilde{w}$, of the Lorentz factor.
Then, one obtains a trial value,
$\tilde{\rho}$, of the rest mass density:
$\tilde{\rho} = \rho_{*}/(\tilde{w} e^{6\phi})$.
\item A trial value, $\tilde{T}$, of the temperature can be
obtained by solving 
\beq
\hat{e} = \hat{e}_{\rm EOS}
(\tilde{\rho}, Y_{e}, \tilde{T}, Y_{\nu e}, Y_{\bar{\nu} e}, Y_{\nu x}),
\eeq 
where $\hat{e}_{\rm EOS}$ is constructed from EOS table.
Note that $\hat{e}$ and $\hat{e}_{\rm EOS}$ in general contain 
contributions from trapped-neutrinos.
One dimensional search over the EOS table is
required to obtain $\tilde{T}$.
\item The next trial value of the Lorentz factor is given by solving 
$\tilde{w} =\sqrt{1+e^{-4\phi}\tilde{\gamma}^{ij}\hat{u}_{i}\hat{u}_{j}\tilde{h}^{-2}}$, 
where the specific enthalpy $\tilde{h}$ is calculated
from EOS table as $\tilde{h} = \tilde{h}(\tilde{\rho}, Y_{e}, \tilde{T})$.
\item Repeat the procedures (i)--(iii) until a required degree of convergence
  is achieved. Convergent solutions of the temperature and $w$ are
  obtained typically within 10 iterations.
\end{enumerate}

On the other hand, in the case where the total lepton fraction 
is evolved, the argument quantities ($\rho, Y_{e}, T$) must be recovered 
from the conserved quantities and $Y_{l}$ 
under the assumption of the $\beta$-equilibrium. 
In this case, two-dimensional reconstruction 
\beq
(Y_{l}, \hat{e}) \ \ \ \Longrightarrow \ \ \ (Y_{e}, T)
\eeq
would be required for a given $\tilde{w}$. 
In this case, there may be in general more than one couple 
of ($Y_{e}$, $T$) which gives the same $Y_{l}$ and $\hat{e}$.
Therefore, I adopt a different method to recover 
($\rho, Y_{e}, T$) \cite{PhD}.

Under the assumption of the $\beta$-equilibrium, the electron fraction
is related to the total lepton fraction as
$Y_{e} = Y_{e}(\rho, Y_{l}, T)$. Using this relation, the original 
EOS table can be reconstructed in terms of the argument quantities 
of ($\rho$, $Y_{l}$, $T$). 
Then, the same strategy as in the above can be adopted. Namely,
\begin{enumerate}
\item Give a trial value, $\tilde{w}$ of $w$. 
Then, one obtains a trial value,
$\tilde{\rho}$, of the rest mass density.
\item A trial value, $\tilde{T}$, of the temperature can be
obtained by solving 
\beq
\hat{e} = \hat{e}_{\rm EOS}(\tilde{\rho}, Y_{l}, \tilde{T}, Y_{\nu x}).
\eeq 
One dimensional search over the EOS table is
required to obtain $\tilde{T}$.
\item The next trial value of $w$ is given by solving 
$\tilde{w} = \sqrt{1+e^{-4\phi}\tilde{\gamma}^{ij}\hat{u}_{i}\hat{u}_{j}\tilde{h}^{-2}}$. 
\item Repeat the procedures (i)--(iii) until a required degree of convergence
  is achieved. The electron and electron neutrino fractions are given as 
  $Y_{e} = Y_{e}(\rho, Y_{l}, T)$, $Y_{\nu e} = Y_{\nu e}(\rho, Y_{l}, T)$,
  $Y_{\bar{\nu} e} = Y_{\bar{\nu} e}(\rho, Y_{l}, T)$ in the new EOS table.
\end{enumerate}
The construction of EOS table in terms of the argument variables of 
($\rho$, $Y_{l}$, $T$) is important in the present implementation.

In the case of a simplified or analytic EOS, 
the Newton-Raphson method may be applied to recover the primitive variables. 
In the case of tabulated EOS, by contrast,
the Newton-Raphson method may not good approach since it requires
derivatives of thermodynamical quantities which in general cannot be 
calculated precisely from tabulated EOS by the finite differentiating 
method (see also Sec. \ref{Sound}).

%%%%%%%%%%%%%%%%%%%%%%
\section{Specific details of microphysics}\label{Details}
%%%%%%%%%%%%%%%%%%%%%%

%%%%%%%%%%%
\subsection{Equation of state}\label{EOS}

{\it Baryons}. While any EOS table can be used in the present code,
an EOS \cite{Shen1,Shen2} based on the relativistic mean field theory
is adopted for baryon EOS (hereafter denoted by Shen EOS) in the present 
version of our code.
Note that the causality is guaranteed to be satisfied in
the relativistic EOS, whereas the sound velocity
sometimes exceeds the velocity of the light in the non-relativistic
framework, e.g., in the EOS by Lattimer and Swesty \cite{LS91}. 
%This is one of the benefits to adopt the relativistic EOS.

{\it Electrons and Positrons}. If a EOS table for baryons   
does not include the contributions of the leptons (electrons, positrons,
and neutrinos if necessary) and
photons, one has to consistently include these contributions to the
table. Electrons and positrons are described as ideal Fermi gases.

To consistently calculate the contribution of the electrons,
the charge neutrality condition $Y_{p} = Y_{e}$ should be solved in
terms of the electron chemical potential
$\mu_{e}$, for each value of the baryon rest-mass density $\rho$ and
the temperature $T$ in the EOS table:
\beq 
n_{e}(\mu_{e},T) \equiv n_{-} - n_{+} = \frac{\rho Y_{p}}{m_{u}}
\label{n_to_mu}
\eeq
in terms of $\mu_{e}$ for given $\rho, T$, and $Y_{p}$.
Here, $m_{u} = 931.49432$ MeV is the atomic mass unit, 
and $n_{-}$ and $n_{+}$ are the total number densities
(i.e., including electron-positron pairs) of
electrons and positrons, respectively. 
Then all other quantities can be calculated from $T$ and $ \mu_{e}$.

{\it Radiations}. 
The contribution of radiations is included in a standard manner:
The radiation pressure and the specific internal energy density are
given by
\beqn
\varepsilon_{r} = \frac{a_{r}T^{4}}{\rho}, \ \ \  
P_{r} = \frac{a_{r}T^{4}}{3},
\eeqn
where $a_{r}$ is the radiation constant
$ a_{r} = (8\pi^{5} k_{B}^{4})(15c^{3}h_{P}^{3})^{-1}$ and $k_{B}$ and $h_{P}$ 
are the Boltzmann's and the Plank's constants respectively.

{\it Trapped-Neutrinos}.
In this paper, the trapped-neutrinos are assumed to interact 
sufficiently frequently with matter that be thermalized. Therefore
they are described as ideal Fermi gases with the matter temperature.
Then, from the neutrino fractions $Y_{\nu}$,
the chemical potentials of neutrinos are calculated by solving
\beq
Y_{\nu} = Y_{\nu}(\mu_{\nu}, T).
\eeq
Using the chemical potentials and the matter temperature, the pressure
and the internal energy of the trapped-neutrinos are calculated.

%%%%%%%%%%%
\subsection{The sound velocity}\label{Sound}

In high-resolution shock-capturing schemes,
it is in general necessary to evaluate the sound velocity $c_{s}$,
\beq
c_{s}^{\,2} = \frac{1}{h}\left[ 
\left.\frac{\partial P}{\partial \rho}\right|_{\epsilon}
+\frac{P}{\rho}
\left.\frac{\partial P}{\partial \epsilon}\right|_{\rho}
\right]. \label{defcs}
\eeq
Here, the derivatives of the pressure are calculated by
\beqn
\left.\frac{\partial P}{\partial \rho}\right|_{\epsilon}
&=&
\sum_{i=B,e,r, \nu}
\left[
  \left.\frac{\partial P_{i}}{\partial \rho}\right|_{T}
  -\left.\frac{\partial P_{i}}{\partial T   }\right|_{\rho}
  \left(
  \sum_{j=B,e,r, \nu}
  \left.\frac{\partial \epsilon_{j}}{\partial \rho}\right|_{T}
  \right)
  \left( \sum_{k=B,e,r, \nu}
  \left.\frac{\partial \epsilon_{k}}{\partial T}\right|_{\rho}
  \right)^{-1}  
\right], \label{Prho} \\
\left.\frac{\partial P}{\partial \epsilon}\right|_{\rho}
&=&
 \left(\sum_{i=B,e,r, \nu}
 \left.\frac{\partial P_{i}}{\partial T}\right|_{\rho}
 \right)
 \left(\sum_{j=B,e,r, \nu}
 \left.\frac{\partial \epsilon_{j}}{\partial T}\right|_{\rho}
 \right)^{-1} , \label{Peps}
\eeqn
where '$B$', '$e$', '$r$', and $\nu$ in the sum denote contributions of 
the baryon, the electrons, radiations,  and neutrinos, respectively.
 
Since there are in general the phase transition regions
in a EOS table for baryons and the EOS moreover may contain 
some non-smooth spiky structures, careful treatments are necessary when 
evaluating the derivatives of thermodynamical quantities. 
In the present EOS table, the derivatives are carefully evaluated 
so that there are no spiky behaviors in the resulting sound velocities. 

%%%%%%%%%%%
\subsection{The local rates}\label{Local}

In this paper,
the electron and positron captures ($\gamma_{\nu}^{\rm ec}$ and
$\gamma_{\nu}^{\rm pc}$) \cite{FFN},
the electron-positron pair annihilation ($\gamma_{\nu}^{\rm pair}$) \cite{CHB86},
the plasmon decays ($\gamma_{\nu}^{\rm plas}$) \cite{RJS96},
and the Bremsstrahlung processes ($\gamma_{\nu}^{\rm Brems}$) \cite{BRT06}
are considered as the local production reactions of neutrinos.
Then, the local rates for lepton fractions are
\beqn
&&\!\! \gamma_{e}^{\rm local} = \gamma_{\nu}^{\rm ec} - \gamma_{\nu}^{\rm pc}, \\
&&\!\! \gamma_{\nu e}^{\rm local} = 
\gamma_{\nu}^{\rm ec} + \gamma_{\nu}^{\rm pair} + \gamma_{\nu}^{\rm plas} +
\gamma_{\nu}^{\rm Brems}, \\
&&\!\! \gamma_{\bar{\nu} e}^{\rm local} = 
\gamma_{\nu}^{\rm pc} + \gamma_{\nu}^{\rm pair} + \gamma_{\nu}^{\rm plas} +
\gamma_{\nu}^{\rm Brems}, \\
&&\!\! \gamma_{\nu x}^{\rm local} = 
\gamma_{\nu}^{\rm pair} + \gamma_{\nu}^{\rm plas} + \gamma_{\nu}^{\rm Brems}.
\eeqn

Similarly, the local energy emission rate $Q_{\nu}^{\rm local}$ is 
the sum of the contributions of the electron and positron captures 
($Q_{\nu}^{\rm ec}$ and $Q_{\nu}^{\rm pc}$) , the electron-positron pair
annihilation ($Q_{\nu}^{\rm pair}$), the plasmon decays 
($Q_{\nu}^{\rm plas}$), and Bremsstrahlung processes ($Q_{\nu}^{\rm Brems}$).

%%%%%%%%%%%
\subsection{The diffusion rates}\label{Diff}

I follow \cite{RL03} for the neutrino
diffusion rates $\gamma_{\nu}^{\rm diff}$ and $Q_{\nu}^{\rm diff}$. 
I present the forms of the diffusion rates in the following for 
convenience. An alternative definition of the diffusion rates 
will be found in \cite{RJS96}.
The cross sections adopted in this paper are those of 
neutrino-nucleus and neutrino-nucleon scattering, 
and neutrino absorptions on free nucleons. 
Explicit forms of these cross sections will be found in \cite{BRT06} 

Ignoring the higher order corrections, the neutrino 
cross sections can be written in general as
\beq
\sigma(E_{\nu}) = E_{\nu}^{2}\tilde{\sigma} , \label{crosssect}
\eeq
where $E_{\nu}$ is the neutrino energy and $\tilde{\sigma}$ is 'cross section'
in which $E_{\nu}^{2}$ dependence is factored out.
Similarly, the opacity and the optical depth are written as
\beqn
&&\!\!\kappa(E_{\nu}) = \sum \kappa_{i}(E_{\nu}) 
= E_{\nu}^{2} \sum \tilde{\kappa}_{i} = E_{\nu}^{2} \tilde{\kappa}, \\
&&\!\!\tau (E_{\nu}) = \int \kappa (E_{\nu}) ds 
= E_{\nu}^{2}\int \tilde{\kappa} ds 
= E_{\nu}^{2} \tilde{\tau}.
\eeqn

Now the neutrino diffusion time may be defined by \cite{RJS96,RL03}
\beq
T_{\nu}^{\rm diff} (E_{\nu}) \equiv
\frac{\Delta x(E_{\nu})}{c}\tau(E_{\nu})
=
E_{\nu}^{2} a^{\rm diff} \frac{\tilde{\tau}^{2}}{c\tilde{\kappa}} 
= E_{\nu}^{2} \tilde{T}_{\nu}^{\rm diff},
\eeq
where the distance parameter $\Delta x (E_{\nu})$ is set to be
\beq
\Delta x (E_{\nu}) = a^{\rm diff} \frac{\tau (E_{\nu})}{\kappa (E_{\nu})}.
\eeq
Here $a_{\rm diff}$ is a parameter which controls the diffusion rates.
In this paper, I adopt $a_{\rm diff} = 3$ as suggested in \cite{RJS96}. 

Finally, the neutrino diffusion rates are defined as
\beqn
&&\!\! N_{\nu}^{\rm diff} \equiv 
\int \frac{n_{\nu}(E_{\nu})}{T_{\nu}^{\rm diff}(E_{\nu})}dE_{\nu}
= \frac{1}{a^{\rm diff}}\frac{4\pi c g_{\nu}}{(h_{P}c)^{3}}
  \frac{\tilde{\kappa}}{\tilde{\tau}^{2}} T F_{0}(\eta_{\nu}), \\
&&\!\! Q_{\nu}^{\rm diff} \equiv 
\int \frac{E_{\nu}n_{\nu}(E_{\nu})}{T_{\nu}^{\rm diff}(E_{\nu})}dE_{\nu}
= \frac{1}{a^{\rm diff}}\frac{4\pi c g_{\nu}}{(h_{P}c)^{3}}
  \frac{\tilde{\kappa}}{\tilde{\tau}^{2}} T^{2} F_{1}(\eta_{\nu}),
\eeqn
from which the diffusion rates 
$Q_{\nu}^{\rm diff}$ and $\gamma_{\nu}^{\rm diff}$ are easily calculated. 
Here, $g_{\nu}$ is the statistical weight factor for neutrinos and 
$ n(E_{\nu})dE_{\nu}$ is the number density of neutrino in 
the range from $E_{\nu}$ to $E_{\nu}+dE_{\nu}$ under the Fermi-Dirac distribution.
%Note that the trapped-neutrinos are assumed to behave as ideal Fermi gases.

%%%%%%%%%%%%%%%%%%%%%%
\section{Validity of general relativistic leakage scheme}\label{Results}
%%%%%%%%%%%%%%%%%%%%%%

\begin{table}[b]
 \begin{center}
  \begin{tabular}{c|c|ccccc} \hline
   Model &  & {\small $\Phi_{c} \le 0.0125 $} & {\small  $  \le \Phi_{c} \le 0.025 $}
           & {\small $  \le \Phi_{c} \le 0.05 $} 
           & {\small $  \le \Phi_{c} \le 0.1 $} & {\small $\Phi_{c} \ge 0.1$} \\
           \hline
  S15    & $\Delta x_{0}$ & 3.26 & 1.60 & 0.820 & 0.414 & 0.217 \\
         & $\eta$  & 1.005 & 1.005 & 1.005 & 1.005 & 1.005  \\
         & $N$     & 444    & 668    & 924    & 1212   & 1532    \\
         & $L$ {\small (km)}& 2330   & 2239   & 2188   & 2124   & 2103    \\ \hline
  S15    & $\Delta x_{0}$ & 5.10 & 2.90 & 1.44 & 0.760 & 0.396 \\
 {\footnotesize low}     & $\eta$  & 1.005 & 1.005 & 1.005 & 1.005 & 1.005  \\
 {\footnotesize resolution}       & $N$     & 316 & 444 & 636 & 828 & 1020  \\
         & $L$ {\small (km)}& 2244 & 2151 & 2073 & 2043 \\ \hline
  \end{tabular}
 \end{center}
\caption{Summary of the regridding procedure. The values of the minimum
 grid spacing $\Delta x_{0}$ (in units of km), 
 the non-uniform-grid factor $\eta$, and
 the grid number $N$ for each range of $\Phi_{c} = 1 -\alpha_c$ are
 listed. }\label{regrid}
\end{table}

\begin{figure}[t]
  \begin{center}
      \includegraphics[scale=1.1]{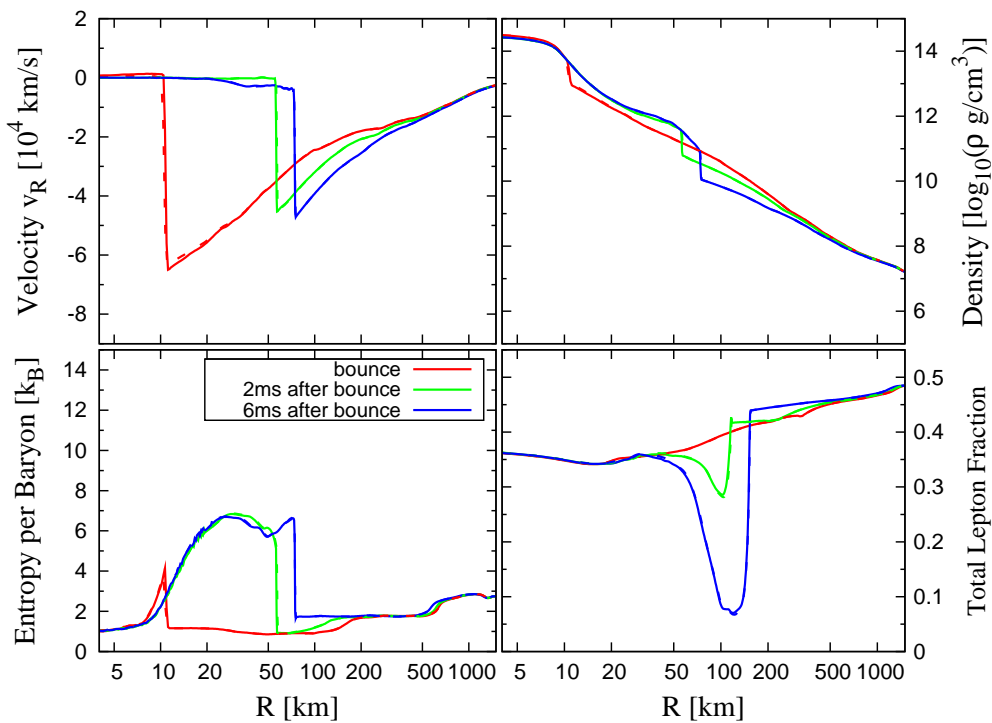}
  \end{center}
  \caption{The radial profiles of the infall velocity, the density, 
    the entropy per baryon and the total lepton fraction at bounce,
    2 ms and 6 ms after bounce. 
    The results for the finer grid resolution (solid curve) and for the coarser grid resolution 
    (the dotted curves) are shown together while they are almost identical.
}\label{qalx}
\end{figure}

\begin{figure}[t]
  \begin{center}
    \includegraphics[scale=1.15]{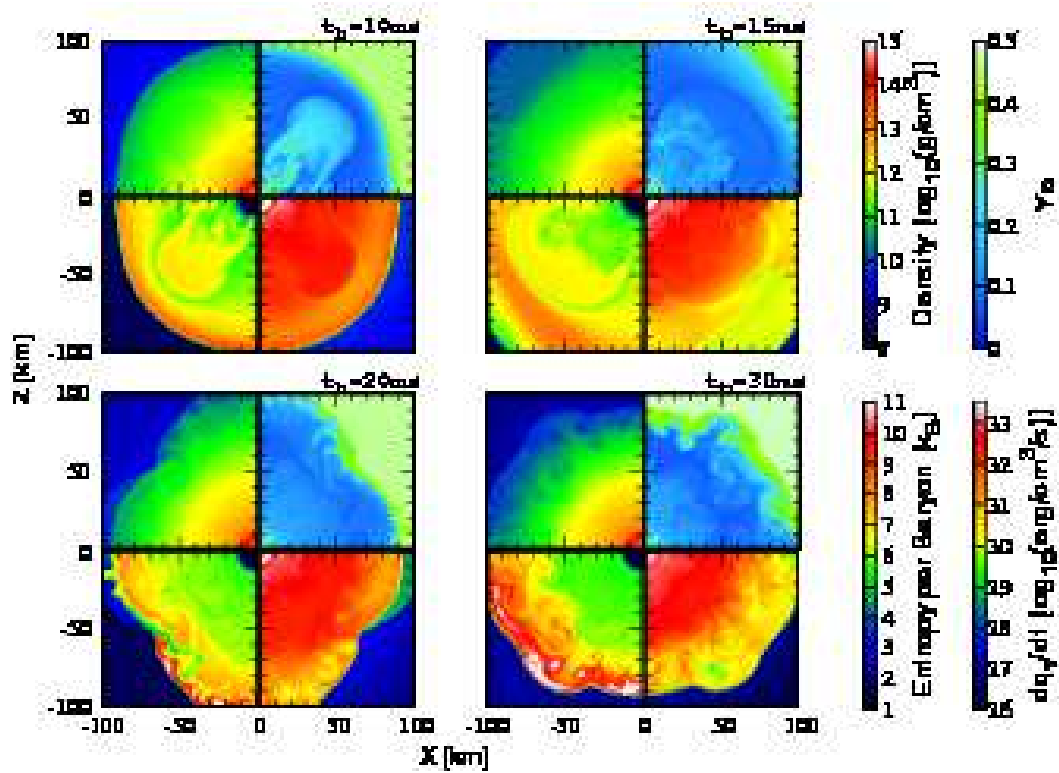}
  \end{center}
  \caption{Snapshots of the contours of the density (top left panels),  
    the electron fraction $Y_{e}$ (top right panels), the entropy per baryon
    (bottom left panels), and the local neutrino energy emission rate 
    (bottom right panels) in the $x$-$z$ plane at selected time slices.  
}\label{con}
\end{figure}

\begin{figure}[t]
  \begin{center}
    \includegraphics[scale=1.0]{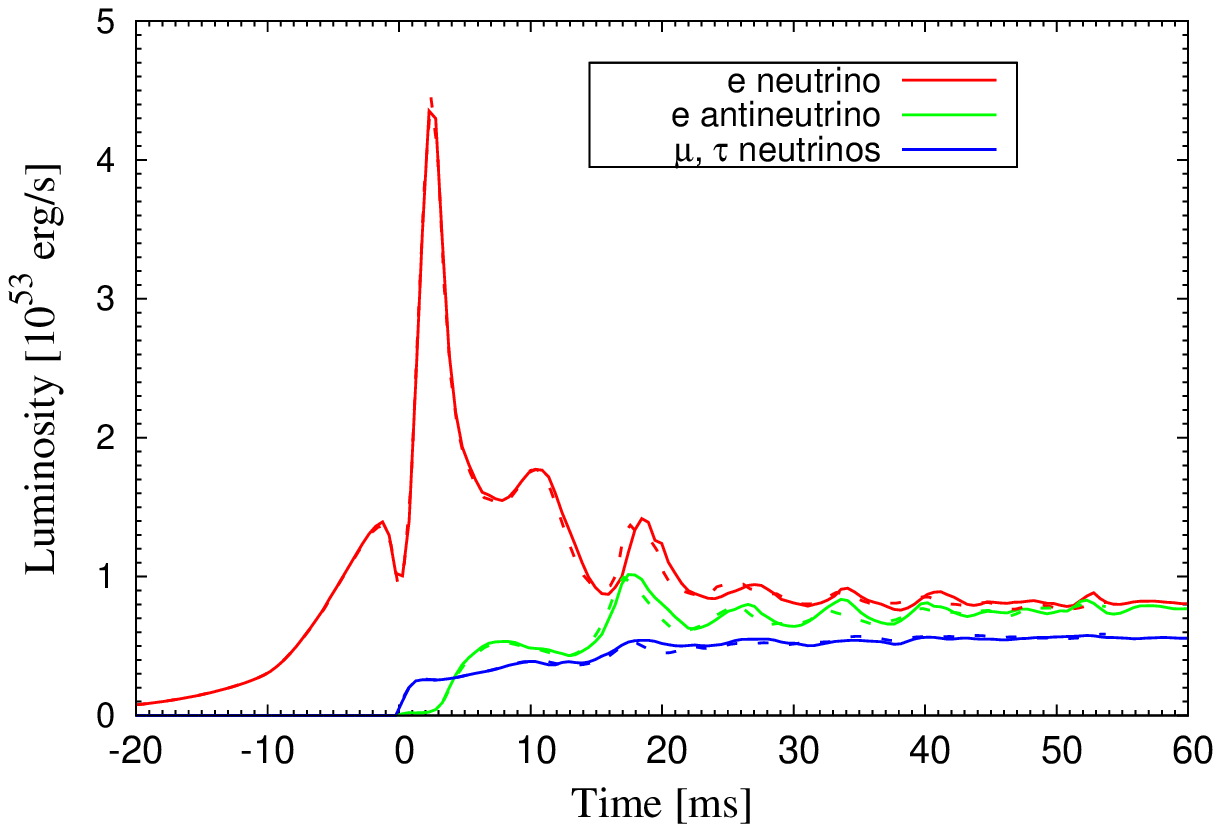}
  \end{center}
  \caption{Time evolution of the neutrino luminosities. 
    The results in the finer grid resolution (solid curves) and 
    in the coarser grid resolution (dashed curves) are shown together. 
    The two results are almost identical until the convective phase sets in,
    while they are not in the convective phase.}\label{nlum}
\end{figure}

\subsection{Brief summary of numerical set up}

The numerical schemes for solving the Einstein's equations are essentially
same as those in \cite{BNS1}; We adopt so-called BSSN 
formulation \cite{BSSN1,BSSN2} and use 4th-order finite difference
scheme in the spatial direction and
the 3rd-order Runge-Kutta scheme in the time integration.
The advection terms such as $\beta^{i}\partial_{i}\phi$ are evaluated
by a 4th-order upwind scheme.
The hydrodynamic equations, the lepton-number conservation equations, 
and equations of streaming-neutrinos are solved using the 
high-resolution centered scheme
\cite{center}.

A nonuniform grid is adopted in the numerical simulation, in which 
the grid spacing increases as
\beq
d x_{j+1} = \eta d x_{j}, \ \ \ \ d z_{l+1} = \eta d z_{l}
\eeq
where $d x_{j} \equiv x_{j+1} - x_{j}$, $d z_{l} \equiv z_{l+1} - z_{l}$ 
and $\eta$ is a constant.
The regridding procedure \cite{Shibata02,Sekiguchi05} is furthermore
used to compute the collapse accurately and to save the CPU
time efficiently. For the regridding, I define an effective
gravitational potential $\Phi_c \equiv 1 -\alpha_c~ (\Phi_c>0)$ where
$\alpha_c$ is the central value of the lapse function. 
In Table \ref{regrid}, parameters of the regridding procedure are
summarized. More detailed set up of the simulation will be found elsewhere 
\cite{YS}. 

As a test problem, I performed a collapse simulation of 
spherical presupernova core.
A presupernova model (S15) of $15M_{\odot}$ with solar metallicity computed in
\cite{WHW02} is adopted as the initial condition.
I follow the dynamical evolution of central part which is composed of the
Fe core and some part of the Si-shell. 
The density, the electron fraction, and the temperature are used to 
calculate other thermodynamical quantities using the EOS table. 

To check the validity of the code, the results are compared with those 
in the state-of-the-art one-dimensional simulations (hereafter, the
reference simulations) in full general 
relativity \cite{Lieb01,Lieb04,Lieb05,Sumi05}, 
where one dimensional general relativistic Boltzmann equation is 
solved for neutrino transfer with relevant weak interaction processes.
Since neutrino heating processes ($\nu_{e}+n \rightarrow p+e^{-}$ and 
$\bar{\nu}_{e}+p\rightarrow n+e^{+}$) are not include in the present 
implementation, and multidimensional effects such as convection cannot be
followed in the one-dimensional reference simulations, 
I pay particular attention in comparing results during the collapse and 
the early phase ($\sim 10$ ms) after the bounce. After that, direct
comparison cannot be done since in the present multidimensional code 
convective activities set in.
As shown below, results in the present simulation and in the 
reference simulations agree well.

%%%%%%%%%%%
\subsection{Comparison of the radial profiles}

The collapse proceeds until the nuclear density is reached in the
central part of the iron core. Then, the inner core experiences the
bounce due to the nuclear repulsive forces, forming a strong shock wave
at the edge of the inner core. 
The shock wave propagates outward and when it crosses the neutrino-sphere, 
spiky burst emissions of neutrinos occur (neutrino bursts):
Neutrinos in hot post-shock region are copiously emitted without
interacting matters.
Eventually, negative gradients of the total lepton fraction are formed behind
the shock since neutrinos carry away the lepton number.
In Fig. \ref{qalx}, we show the radial profiles of the infall velocity, 
the density, the entropy per baryon and the total lepton fraction at
selected time slices. 

The results agree at least semi-quantitatively with those in
\cite{Lieb01,Lieb04,Lieb05,Sumi05}. In particular, the radial profiles of 
the infall velocity, the density, and the entropy per baryon show good agreements.
No such good agreements was reported in the previous Newtonian simulations
in which leakage schemes are adopted \cite{leak1,leak2,leak3,leak4,leak5}. 
The negative gradients quantitatively are little bit steeper in the
present simulation. The reason may be partly because the {\it transfers} of
lepton-number and energy are not solved in the present leakage scheme.
Except for this quantitative difference, the two results agree well.
It is found that the difference can be reduced by adjusting the 
parameter $a_{\rm diff}$ introduced in Sec. \ref{Diff}.

Recall that regions of negative $Y_{l}$ gradient are known to be
convectively unstable \cite{Bethe1990}. Convective activities indeed set
in in the present simulation as shown in Fig. \ref{con}.

%%%%%%%%%%%
\subsection{Comparison of the Neutrino luminosities}

Comparisons of the neutrino luminosities are particularly important
since they depend on both implementations of weak interactions 
(especially electron capture in the present case) and treatments of
neutrino cooling (the detailed leakage scheme). 
Also, accurate estimations of neutrino luminosities 
would be primarily important for astrophysical applications,
since neutrinos carry away the most of energy liberated during the
collapse as the main cooling source.

In Fig. \ref{nlum}, I show neutrino luminosities calculated 
according to \cite{SSR07}
\beq
L_{\nu} = \int \alpha e^{6\phi} u_{t} \dot{Q}^{\rm leak}_{\nu} d^{3}x ,
\eeq
as functions of $t-t_{\rm bounce}$ where $t_{\rm bounce}$ is time at 
the bounce. The result also agrees approximately with that in the
reference simulations.
The neutrino bursts occur when the shock wave crosses the
neutrino-sphere soon after the bounce.
The peak luminosity at the neutrino burst is 
$L_{\nu _{e}} \approx 4.5 \times 10^{53}$ ergs/s in the present
simulation, which agrees well with that in the reference simulations.
The peak luminosity and the duration (width) of the neutrino burst
emission can be improved by adjusting the parameter $a_{\rm diff}$.
The modulation found in the later phase $t-t_{\rm bounce} > 10$ ms is due to
convective activities driven by negative gradients of electron fraction
and entropy per baryon.

Thus, Fig. \ref{nlum} illustrates that the present detailed leakage
scheme works fairly well and may be applied to simulations of rotating
core collapse to a black hole and mergers of binary neutron stars.

In the previous simulations based on the leakage scheme \cite{leak4,PhD}
where the single 'neutrino-trapping' density is adopted, the
luminosities do not agree with that in the reference simulations.
In particular, the luminosities at the neutrino bursts are quite
different.

%%%%%%%%%%%
\subsection{Convergence}

In Figs. \ref{qalx} and \ref{nlum}, I show results in the higher resolution
(solid curves) and the lower resolution (dashed curves).  
The radial profiles of the two resolutions are almost identical, showing
that convergent results are obtained in the present simulation 
(see Fig. \ref{qalx}).
In the time evolution of neutrino luminosities (see Fig. \ref{nlum}),
the two results are almost identical before the convective activities set in.
In the later phase, on the other hand, the two results shows slight difference.
Since the convection and the turbulence can occurs in a infinitesimal
scale length, the smaller-scale convection and turbulence are captured
in the finer grid resolution. Further discussions associated with the
convergence and numerical accuracy will be found in \cite{YS}.

%%%%%%%%%%%%%%%%%%%%%%
\section{Summary and Discussions}\label{Summary}
%%%%%%%%%%%%%%%%%%%%%%

\subsection{Summary}

In this paper, I presented an implementation of the weak interactions and
the neutrino cooling in the framework of full general relativity.
Since the characteristic timescale of weak interaction processes
$t_{\rm wp} \sim \vert Y_{e}/\dot{Y}_{e} \vert$ 
is much shorter than the dynamical timescale $t_{\rm dyn}$ in hot dense matters, 
stiff source terms appears in the equations.
In general, an implicit scheme may be required to solve them \cite{Bruenn85}.
However, it is not clear whether implicit schemes do work or not in the 
relativistic framework.
The Lorentz factor is coupled with the rest mass density and the energy 
density. The specific enthalpy is also coupled with the momentum. 
Due to these couplings, it is very complicated to recover the primitive 
variables and the Lorentz factor from conserved quantities. 
Therefore I proposed an explicit method to solve the equations noting
that the characteristic timescale of neutrino leakage from the system
$t_{\rm leak}$  is much longer than $t_{\rm wp}$ and is comparable to
$t_{\rm dyn}$.

By decomposing the energy tensor of neutrino into the trapped-neutrino
and the streaming-neutrino parts, the equations for
the energy momentum tensor can be rewritten so that the source terms are characterized 
by the leakage timescale $t_{\rm leak}$ (see Eqs. (\ref{T_Eq_M}) and (\ref{T_Eq_nuS})).
The lepton-number conservation equations, on the other hand, include the source terms
characterized by the WP timescale. Therefore the {\it limiters} for the stiff source terms 
are introduced to solve the lepton-number conservation equations explicitly 
(see Sec. \ref{Lepton}).

In the numerical relativistic hydrodynamics, it is required to
calculate the primitive variables and the Lorentz factor 
from the conserved quantities.
In this paper, I develop a robust and stable procedure for it 
(Sec. \ref{Reconst}).

Finally, to check the validity of the present implementation, I performed 
a collapse simulation of spherical presupernova core and compared the
results with those obtained in the state-of-the-art one-dimensional 
simulations in full general relativity \cite{Lieb01,Lieb04,Lieb05,Sumi05}.
As shown in this paper, results in this paper agree well with those 
in the state-of-the-art simulations.
Thus the present implementation will be applied to simulations of rotating core collapse 
to a black hole and mergers of binary neutron stars.

\subsection{Discussions}

Since the present implementation of the microphysics is simple and
explicit, it has advantage that the individual microphysical processes
can be easily improved and sophisticated.

For example, the neutrino emission via the electron capture can be
easily sophisticated as follows.
To precisely calculate the electron capture rate, 
the complete information of the parent and daughter nuclei are required.
In the nuclear equations of state currently available, however, 
a representative single-nucleus average for the true ensemble of
heavy nuclei is adopted. The representative is usually the most 
abundant nuclei.
The problem in evaluating the capture rate is that the nuclei
which cause the largest changes in $Y_{e}$ are neither the most
abundant nuclei nor the nuclei with the largest rates, but the
combination of the two. In fact, the most abundant nuclei tend to have
small rates since they are more stable than others,
and the fraction of the most reactive nuclei tend to be small \cite{AFWH94,Janka06}.
Assuming that the nuclear statistical equilibrium (NSE) is achieved, the
electron capture rates under the NSE ensemble of heavy
nuclei may be calculated for given ($\rho$, $Y_{e}$, $T$).
Such a numerical rate table can be easily employed in the present
implementation.

Also, the neutrino cross sections can be improved.
As summarized in \cite{Horowitz}, there are a lot of higher order 
corrections to the neutrino opacities.
Note that small changes in the opacities 
may result in much larger changes in the neutrino luminosities, since the 
neutrino energy emission rates strongly depend on the temperature 
and the temperature at the last scattering surface 
($\tau_{\nu}\sim \sigma T^{2} \sim 1$) changes as $T \sim \sigma^{-1/2}$.
Although the correction terms are in general very complicated, it is 
straightforward to include the corrections in the present
implementation. Note that the corrections become more important for
higher neutrino energies. Therefore, the correction terms might play
roles in the collapse of population III stellar core and the formation
of a black hole in which very high temperatures ($T>100$ MeV) are achieved.
I already started studies to explore the importance of these corrections
in the case of black hole formation.

As briefly described in the introduction, one of main drawbacks of the
present implementation of the neutrino cooling is that the {\it
transfer} of neutrinos are not solved.
Although to {\it fully} solve the transfer equations of neutrinos is far beyond
the scope of this paper, there are a lot of rooms for improvements in
the treatment of the neutrino cooling. For example, the relativistic
moment formalism \cite{AS72,Thorne81}, in particular the so-called M1 closure
formalism, may be adopted.
For this purpose, a more sophisticated treatment of the closure relation
for $P_{\alpha \beta}$ is required.
For example, one may adopt the Eddington tensor of the form \cite{Levermore}
\beq
P_{\alpha \beta} = \left[
\frac{1-\chi}{2}\gamma_{\alpha \beta}  
+ \frac{3\chi-1}{2}\frac{F_{\alpha}F_{\beta}}{F_{\gamma}F^{\gamma}} \right]E, 
\eeq
where $\chi$ is the (variable) Eddington factor.
In the diffusion limit where the neutrino pressure is isotropic $\chi=1/3$,
while in the free streaming limit $\chi=1$.
We plan to implement a relativistic M1 closure formalism for the
neutrino transfer in the near future.

To conclude, the present implementation of microphysics in full general
relativity works fairly well. We are now in the standpoint where simulations
of stellar core collapse to a black hole and merger of compact stellar binaries 
can be performed including microphysical processes. 
Fruitful scientific results will be reported in the near future.

%%%%%%%%%%%%%%%%%%%%%%
\section*{acknowledgments}
%%%%%%%%%%%%%%%%%%%%%%

I thank M. Shibata and L. Rezzolla for valuable discussions, and the
referees for valuable comments. 
I also thank T. Shiromizu and T. Fukushige for their 
grateful aids.
Numerical computations were performed on the NEC SX-9 at the data
analysis center of NAOJ and on the NEC SX-8 at YITP in Kyoto University. 
This work is partly
supported by the Grant-in-Aid of the Japanese Ministry of Education, Science,
Culture, and Sport (21018008,21105511).

%%%%%%%%%%%%%%%%%%%%%%

\end{document}